\documentclass{article}

\usepackage{microtype}
\usepackage{graphicx}
\usepackage{subfigure}
\usepackage{placeins}
\usepackage{booktabs} 

\usepackage{hyperref}

\usepackage[accepted]{icml2020}

\icmltitlerunning{Rain Sensing Automatic Car Wiper Using AT89C51 Microcontroller}

\begin{document}

\twocolumn[
\icmltitle{Rain Sensing Automatic Car Wiper Using AT89C51 Microcontroller}



\icmlsetsymbol{equal}{*}

\begin{icmlauthorlist}
\icmlauthor{Abhishek Das}{DJSCE}
\icmlauthor{Aditya Desai}{DJSCE}
\icmlauthor{Vivek Dhuri}{DJSCE}

\end{icmlauthorlist}

\icmlaffiliation{DJSCE}{UG. Student, Department of Electronics and Telecommunication, DJSCE, Mumbai, India}
\icmlaffiliation{DJSCE}{UG. Student, Department of Electronics and Telecommunication, DJSCE, Mumbai, India}
\icmlaffiliation{DJSCE}{UG. Student, Department of Electronics and Telecommunication, DJSCE, Mumbai, India}



\vskip 0.3in
]



\printAffiliationsAndNotice{}  


\section{Abstract}
\label{submission}
The turn of the century has seen a tremendous rise in technological advances in the field of automobiles. With 5G technology on its way and the development in the IoT sector, cars will start interacting with each other using V2V communications and become much more autonomous. In this project, an effort is made to move in the same direction by proposing a model for an automatic car wiper system that operates on sensing rain and snow on the windshield of a car. We develop a prototype for our idea by integrating a servo motor and raindrop sensor with an AT89C51 Microcontroller.

\section{Introduction}
\label{submission}

\begin{figure}
\vskip 0.2in
\begin{center}
\centerline{\includegraphics[width=\columnwidth,height= 6 cm]{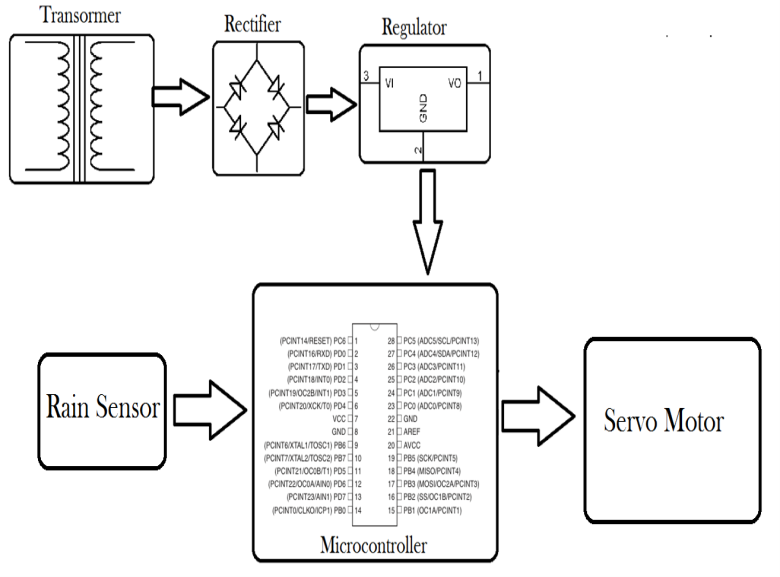}}
\caption{Block Diagram of Proposed Model}
\label{BlockDiagram}
\end{center}
\vskip -0.4in
\end{figure}

Today’s car wipers are manual systems that work on the principle of manual switching. So here we propose an automatic wiper system that automatically switches ON on detecting rain and stops when the rain stops. Our project brings forward this system to automate the wiper system not need manual intervention. For this purpose, we use a rain sensor along with a microcontroller to drive the wiper motor. Our system uses a rain sensor to detect rain, this signal is then processed by a microcontroller to take the desired action. The rain sensor works on the principle of using water for completing its circuit, so when rain falls on it, the circuit gets completed and sends out a signal to the microcontroller. The microcontroller now processes this data and controls the motor. This system is equally useful for Aircraft and a smaller version of this can be used by motorbikers in their helmets so that they can drive easily in rains. Figure \ref{BlockDiagram} shows the block diagram for our proposed idea.

We use Assembly Language Coding using arm Keil \( \mu \)Vision 5 interface. For PCB designing, we use EAGLE software \cite{monk2017make} and for circuit simulation, we have used Proteus Design Suite \cite{su2010application}.

In what follows, we discuss the related prior work for such a problem in the next section (3), followed by defining the problem statement (4) and discussing our novel approaches in section (5). We then present our Experimental setup in section (6) followed with its results and discussion in section (7). Finally, we end the discussion with conclusion and future directions in the last section (8).
The scripts and circuit designs are publicly available \href{https://github.com/Abhishek0697/Rain-Sensing-Automatic-Car-Wiper-using-AT89C51-Microcontroller/tree/master/}{here}.

\section{Related Work}

In the current scenario, only high-end vehicles employ intelligent rain-sensing automatic wiper systems. Our system is modeled to demonstrate how useful is an automatic wiper system that adjusts speed itself based on rainfall intensity. Such a system improves the safety of a ride. There are many instances of accidents occurring during heavy rainfall due to lack of proper vision. In many cases, these accidents were due to manual errors (for example: not increasing the speed of the wiper) from the driver. An automatic, intelligent system like ours removes any manual errors. Our system adjusts wiper speed according to the intensity of rainfall and hence improves safety. Nowadays some models of Ford and Hyundai are also implementing an automatic wiper system in their vehicles[1].

\begin{figure}[h]
\vskip 0.2in
\begin{center}
\centerline{\includegraphics[width=\columnwidth,height= 8 cm]{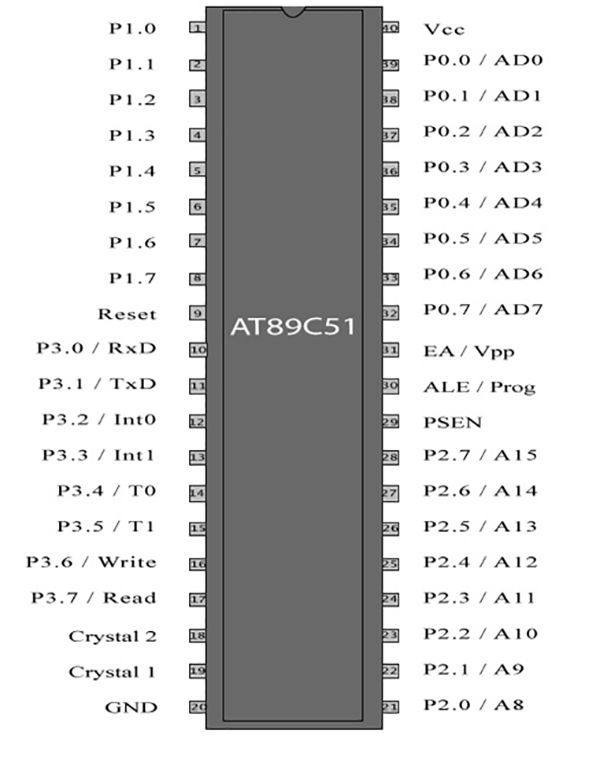}}
\caption{8051 Microcontroller pin configuration}
\label{pinconfig}
\end{center}
\vskip -0.4in
\end{figure}

Now, we discuss some prior work in this area. Since the rain sensor used in the automatic wiper system is expensive \cite{kulkarnisemi} designed a semi-automatic rain wiper that could be installed in economic vehicles. Their semi-automatic rain wiper had Cup sensor which was based on the principle of rate of water flow and volume of water. The cup sensor was made up of a conical cup with probes at different levels of height. These levels of the probe were used to increase the wiper speed. Therefore, depending on the rain intensity the wiper system could change the speed. Their design was economical and had three different stages of rain intensity.

Similarly, \cite{ashik2014automatic} designed automatic wipers with mist control that worked with three different rain intensities, which are drizzling, medium rain and heavy rain. The automatic wiper and internal wiper uses the combination of a sensor, microcontroller and the wiper motor. The external sensor and internal sensor is based on the principle of conductance. The microcontroller actuates the speed of the wiper motor by measuring the rain intensity as detected by the external motor. Similarly, the internal mist controllers are placed on the windshield which detects the mist signalling the controller to actuate the internal wiper motor.

\section{Experimental Setup}

\subsection{AT89C51 Microcontroller}
The AT89C51 \cite{mazidi20058051} is a low-power, high-performance CMOS 8-bit microcomputer with 4K bytes of Flash programmable and erasable read-only memory.  The on-chip Flash allows the program memory to be reprogrammed in-system or by a conventional non-volatile memory programmer. By combining a versatile 8-bit CPU with Flash on a monolithic chip, the Atmel AT89C51 is a powerful microcomputer that provides a highly-flexible and cost-effective solution to many embedded control applications. Figure \ref{pinconfig} shows the pin configuration of 8051 Microcontroller.

\begin{figure}
\vskip 0.2in
\begin{center}
\centerline{\includegraphics[width=4cm, height=4cm]{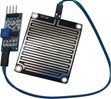}}
\caption{Rain Sensor Module}
\label{RainSensorModule}
\end{center}
\vskip -0.5in
\end{figure}

\subsection{Rain Sensor Module}
A rain sensor module is an easy tool for rain detection \cite{guptadesign}. It can be used as a switch when a raindrop falls through the raining board and for measuring rainfall intensity. Figure \ref{RainSensorModule} shows a depiction of a typical Rain Sensor Module. Due to its compact design and light weight, it can be easily attached into any system. The module features, a rain board, and the control board that is separate for more convenience, a power indicator LED, and sensitivity adjustable through a potentiometer. A raindrop sensor is a board coated with nickel in the form of lines. It works on the principle of ohms law. When there is no raindrop on board. Resistance is high so we get high voltage according to V=IR. When raindrop present it reduces the resistance because water is a conductor of electricity and the presence of water connects nickel lines in parallel so reduced resistance and the reduced voltage drop across it.

\begin{figure}
\vskip 0.2in
\begin{center}
\centerline{\includegraphics[width=4cm, height=4cm]{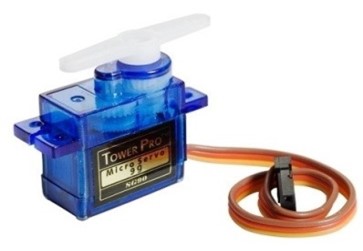}}
\caption{Servo Motor}
\label{Servo_Motor}
\end{center}
\vskip -0.6in
\end{figure}

\begin{figure}
\vskip 0.2in
\begin{center}
\centerline{\includegraphics[width=\columnwidth]{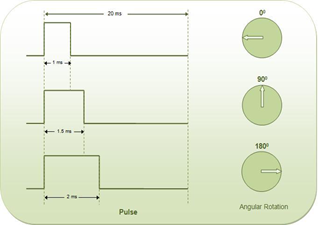}}
\caption{Operation of Servo based on Pulse Width Modulation }
\label{Servooperation}
\end{center}
\vskip -0.4in
\end{figure}

\subsection{Servo Motor}
Servo motors \cite{sachin2013design} are self-contained mechanical devices that are used to control the machines with great precision. . Usually the servo motor is used to control the angular motion from 0° to 180° and 0° to 90°. The servo motor can be moved to a desired angular position by sending Pulse Width Modulated \cite{holtz1992pulsewidth} signals on the control wire. The servo understands the language of pulse position modulation. A pulse of width varying from 1 millisecond to 2 milliseconds in a repeated time frame is sent to the servo around 50 times in a second. The width of the pulse determines the angular position. 
For example, a pulse of 1 millisecond moves the servo towards 0°, while a 2 milliseconds wide pulse would take it to 180°. The pulse width for in-between angular positions can be interpolated accordingly. Thus a pulse of width 1.5 milliseconds will shift the servo to 90°. It must be noted that these values are only approximations. The actual behavior of the servos differs based on their manufacturer. A sequence of such pulses (50 in one second) is required to be passed to the servo to sustain a particular angular position. When the servo receives a pulse, it can retain the corresponding angular position for the next 20 milliseconds. So a pulse in every 20 millisecond time frame must be fed to the servo. Figure \ref{Servo_Motor} shows an example of the servo motor we have used in our implementation, while Figure \ref{Servooperation} shows the operation of servo motor based on Pulse Width Modulated signals.

\begin{figure}[t]
\vskip 0.2in
\begin{center}
\centerline{\includegraphics[width=\columnwidth]{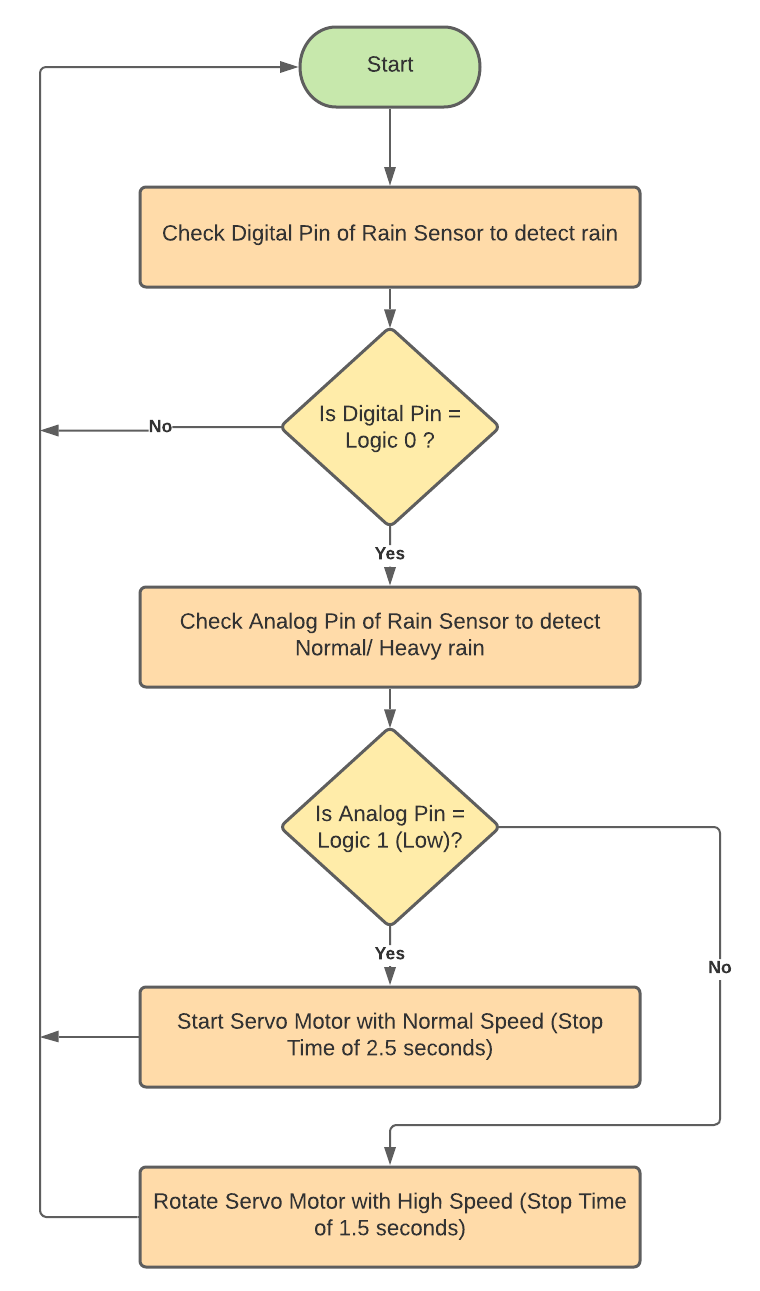}}
\caption{Flowchart}
\label{Flowchart}
\end{center}
\vskip -0.2in
\end{figure}

\begin{figure}
\vskip 0.2in
\centerline{\includegraphics[width=\columnwidth, height =8 cm]{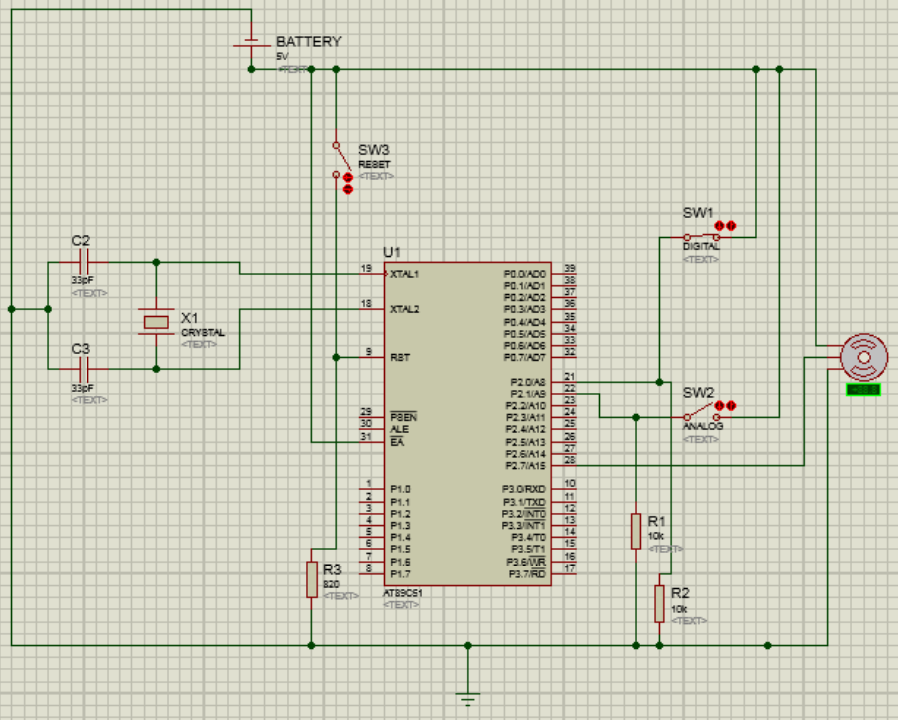}}
\caption{PROTEUS Simulation Diagram}
\label{PROTEUS_Simulation_Diagram}
\vskip -0.2in
\end{figure}

\begin{figure}[t]
\vskip 0.2in
\begin{center}
\centerline{\includegraphics[width=\columnwidth, height = 8 cm]{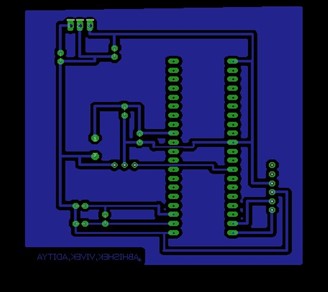}}
\caption{Layout of our Printed Circuit Board in EAGLE software }
\label{LayoutofPCB}
\end{center}
\vskip -0.4in
\end{figure}

\subsection{Circuit Simulation and PCB Designing}
\begin{figure}[h!]
\vskip 0.2in
\begin{center}
\centerline{\includegraphics[width=\columnwidth, height = 8 cm]{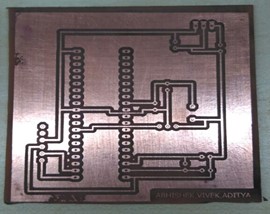}}
\caption{Printed Circuit Board after Etching process}
\label{EtchedPCB}
\end{center}
\vskip -0.4in
\end{figure}

Proteus Design Suite by Labcenter Electronics provides a simple interface to design and simulate various circuits. It has a variety of electronic components and settings for each of them to choose from and is an efficient method to test the initial circuits for sanity checks before implementation. It has the option of adding various switches and connect and visualize the flow in real-time, providing error logs and failure cases. Figure \ref{PROTEUS_Simulation_Diagram} demonstrates the simulation of our circuit design. After selecting all the components and verifying the simulation on the software, we then start implementing the actual PCB designing process which includes steps like printing a layout from EAGLE PCB design software, etching the PCB, drilling, integrating and soldering all the components and finally testing the prototype. Figure \ref{LayoutofPCB} shows the Layout of our printed circuit board on the EAGLE Software and Figure \ref{EtchedPCB} shows the board after etching process. 

\section{Results and Conclusion}

\begin{figure}[h!]
\vskip 0.2in
\begin{center}
\centerline{\includegraphics[width=\columnwidth]{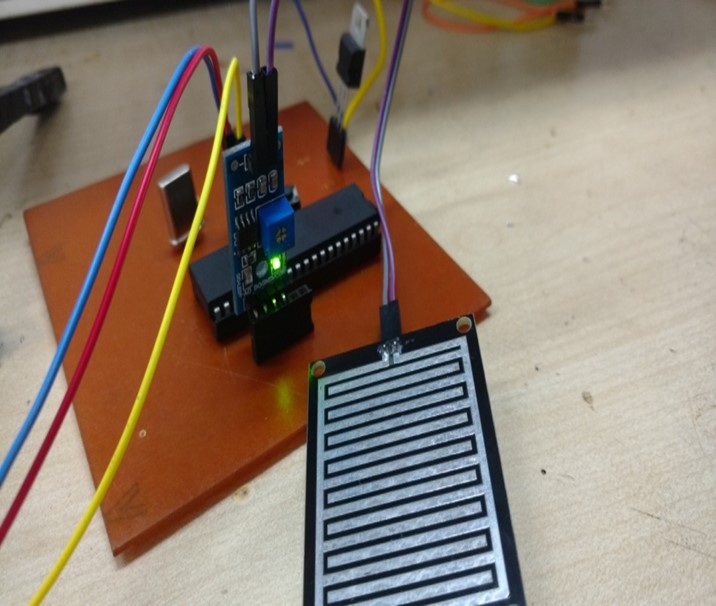}}
\caption{Working prototype}
\label{Model}
\end{center}
\vskip -0.3in
\end{figure}

Thus, we have implemented a model that senses rains and automatically switches on the wiper and adjusts its speed according to the intensity of the rain. As the intensity of the rain increases, the speed of the wiper increases to a certain level. Figure \ref{Flowchart} shows the workflow for our proposal. The microcontroller checks for the digital pin and analog pin inputs of the rain sensor. When there is slight water on the sensor, the digital pin is set to logic '0'. This is used to detect presence of rain water. To check the intensity of rain, we monitor the analog pin output of the rain sensor, whose threshold can be adjusted manually through an attached Potentiometer to indicate how much water should be considered as high rain. According to our observations, the wiper takes 2.2 seconds when a drop of water is poured on the sensor, while it takes only 1.4 seconds when the sensor is submerged in a glass of water. We learned how to interface servo motor with AT89C51 Microcontroller and the rain sensor module interfacing with AT89C51 Microcontroller. Figure \ref{Model} shows the prototype we have developed to demonstrate our idea.

\section{Future Scope and Market Potential}

The world will one day move in self-driving cars is already evident in a series of functions that today’s cars have begun to perform without human intervention. Even in the models sold in India, some cars tell you the route and journey time, park on their own, start the wipers if it is raining, switch on the lights if it gets dark, warn you of moving objects at night, inform you of a school in the vicinity, detect dangerous lane departures and raise alarm if the driver is drowsy. You can control music volume by moving fingers in the air and telling your car what song to play.

One of the primary objectives in this task was to make a design which is compact and easy to integrate with a complex system such as a vehicle. Also, we wanted to demonstrate how these relatively novel sensors can be integrated with a microcontroller to develop an application. Modifications in the circuit can be made with the objective of creating a system on-chip, which can be easily plugged into existing vehicles. The sensor proposed in this model is low cost and efficient to a great extent, however with the development of more high quality and accurate sensors, much more desirable and reliable outputs can be obtained. Another interesting area to explore into is  controlling the speed of the wiper to a more accurate sense. Currently, the wiper moves at two different speeds. By modifying the code, we can have different speeds for a different amount of rain. Also, we can use this automated car wiper along with other automated features to make a Smart Car.

\FloatBarrier

\bibliography{example_paper}
\bibliographystyle{icml2020}

\end{document}